\documentstyle[11pt,appb]{article} 

\begin{document}
\title{CHIRAL SYMMETRY IN MATTER\thanks{Talk presented at the NATO Advanced 
Research Workshop on the Structure of Mesons, Baryons and Nuclei, Krakow, May,
1998}}
\author{Michael C. Birse
\address{Theoretical Physics Group, Department of Physics and Astronomy,\\
University of Manchester, Machester, M13 9PL, UK}}
\maketitle
\begin{abstract}
Soft-pion theorems are used to show how chiral symmetry constrains the
contributions of low-momentum pions to the quark condensate, the pion decay
constant and hadron masses, all of which have been proposed as signals of
partial restoration of chiral symmetry in matter. These have contributions of
order $T^2$ for a pion gas or of order $m_\pi$ for cold nuclear matter, which
have different coefficients in all three cases, showing that there are no
simple relations between the changes to these quantities in matter. In
particular, such contributions are absent from the masses of vector mesons and
nucleons and so these masses cannot scale as any simple function of the quark
condensate. More generally, pieces of the quark condensate that arise from
low-momentum pions should not be associated with partial restoration of chiral
symmetry.
\end{abstract}
  
\section{Introduction}

One of the main topics being addressed by this workshop is the partial
restoration of chiral symmetry in nuclear or hadronic matter and the
possibility that signals of this restoration have been seen in relativistic
heavy-ion collisions. Although a rather unified picture seemed to be emerging
 from some of the previous contributions, what I want to do here is to unravel
some of the threads in this picture by examining whether changes in quantities
like the quark condensate, the pion decay constant and hadron masses are in
fact related to each other, and the extent to which they can be interpreted in
terms of symmetry restoration.

Central to these questions are the soft-pion theorems that embody the
consequences of chiral symmetry for interactions of low-momentum
pions~\cite{dgh}. The focus of this contribution will be the constraints that
these place on the properties of hadrons in matter. I will devote most of my
discussion to the case of a warm gas of pions, where the calculations are
somewhat cleaner than for nuclear matter. In fact the role of chiral symmetry
in the pion gas is rather an old topic~[2--12] and it is rather embarrassing to
admit how long it has taken me learn the lessons from it. However I want to
stress that these lessons are general ones and apply equally to the case of
cold nuclear matter~[13--16].

The chiral isospin symmetry SU(2)$\times$SU(2) is a good approximate symmetry
of the QCD Lagrangian, broken only by the small current masses of the up and
down quarks. This symmetry is realised in the hidden (``spontaneously broken")
mode since the QCD vacuum is not chirally invariant. This can be pictured in
terms of a Mexican-hat potential for the vacuum with a ``chiral circle" of
degenerate vacuum states running round its brim. The pions are, approximately,
the corresponding massless Goldstone bosons. The nonzero quark condensate (the
scalar density of quarks in the vacuum $\langle\overline\psi\psi\rangle$) is
an example of an order parameter for the hidden symmetry. Other indications of
the hidden nature of this symmetry include the nonzero pion decay constant
$f_\pi$, the absence of degenerate parity doublets in the hadronic spectrum,
and the masses of constituent quarks.

In dense hadronic matter, which could be either a gas of pions or nuclear 
matter, we expect the vacuum to move towards the phase with manifest chiral 
symmetry. Signals of such a partial restoration of the symmetry that have been
suggested include:
\begin{itemize}
\item a decrease in the magnitude of the quark condensate,
\item a smaller pion decay constant, 
\item smaller splittings between states of opposite parity in the hadronic 
spectrum, for example the $\rho$ and $a_1$ mesons 
\item decreased hadron masses.
\end{itemize}
Although much attention has focussed on the last of these, this effect has
been demonstrated only in mean-field approaches such as the linear sigma and
Nambu--Jona-Lasinio models. The general arguments outlined here show that
pionic fluctuations make significant, but very different, contributions to
these quantities and so it is important to go beyond the mean-field
approximation in studying hadron properties in matter.

\section{Soft-pion theorems}

The basic tools for elucidating the consequences of chiral symmetry for the
interactions of pions are the ``soft-pion theorems"~\cite{dgh}. These are
obtained from physical matrix elements involving pions with zero 3-momentum by
extrapolating them to zero energy. This (off-shell) extrapolation is done
using the PCAC pion field
\begin{equation}
\mbox{\boldmath $\phi$}={\partial_\mu {\bf A}^\mu\over f_\pi m_\pi^2},
\end{equation}
which connects pions to chiral symmetry. For example, the soft-pion limit of
the pion propagator defined using this field gives the
Gell-Mann--Oakes--Renner (GOR) relation,
\begin{equation}\label{GOR}
\bar m\langle 0|\overline\psi\psi|0\rangle\simeq -m_\pi^2 f_\pi^2,
\end{equation}
which relates the pion mass to the expectation value of the explicit symmetry
breaking piece of the QCD Hamiltonian, ${\cal H}_{\rm SB}=\bar m\overline\psi
\psi$. Taking a typical value of $\bar m\simeq 7$ MeV for the average of the
up- and down-quark current masses, we find that the quark condensate in
vacuum is $\langle\overline\psi\psi\rangle\sim -3$ fm$^{-3}$.

For a more general matrix element we have
\begin{equation}
\langle \alpha|{\cal O}|\beta\pi(q)\rangle\simeq -{i\over f_\pi}
\langle \alpha|[Q_5,{\cal O}]|\beta\rangle,
\end{equation}
up to corrections involving the pion momentum $q$ or the explicit symmetry
breaking strength $m_\pi^2$. The connection between a commutator with an axial
charge operator and creation or annihilation of a pion shows that a
low-momentum pion can be thought of as acting like a chiral rotation. This
implies that any pion scattering amplitude should vanish as $q\rightarrow 0$
in the chiral limit ($m_\pi^2=0$). A central role in the discussion here will
be played by the isospin-averaged amplitude for scattering of a pion from some
other hadron. The leading terms in the chiral expansion of this amplitude are 
of order $q^2$ and $m_\pi^2$.

In studying changes to the quark condensate, we shall need the scalar densities
of quarks in hadrons. For the pion this can be evaluated with the help of a
soft-pion theorem:
\begin{equation}\label{sigpipi}
\langle\pi|\bar m\overline\psi\psi|\pi\rangle\simeq-{1\over f_\pi^2}
\langle 0|[Q_5,[Q_5,{\cal H}_{\rm SB}]]|0\rangle
\simeq m_\pi^2, 
\end{equation}
where the GOR relation (\ref{GOR}) has been used to express the matrix element
in terms of $m_\pi^2$. Factoring out the covariant normalisation of $2m_\pi$,
we are left with the pion-pion sigma commutator $\sigma_{\pi\pi}=m_\pi/2=70$
MeV. This corresponds to an integrated scalar density of quarks in a pion of
$\sigma_{\pi\pi}/\bar m\sim 10$. This large number shows that pions make
important contributions to the scalar density of quarks in hadrons or matter.
These contributions are proportional to the scalar density of pions in the
state of interest, $\langle\alpha|{1\over 2}\mbox{\boldmath
$\phi$}^2|\alpha\rangle$.

The corresponding matrix element for a nucleon can be found from the 
pion-nucleon sigma commutator, $\sigma_{\pi{\scriptscriptstyle N}}
=\langle N|\bar m\overline\psi\psi|N\rangle\simeq 45\pm 7$ MeV~\cite{gls}.
This corresponds to an integrated scalar density of quarks in a nucleon of
$\sim 6$. In contrast, simple relativistic quark models would give values of
2--3. The difference arises from the pion cloud of the nucleon, which
typically contributes about 25 MeV to $\sigma_{\pi{\scriptscriptstyle N}}$ in
chiral bag or soliton models~\cite{cbsm}.

\section{Warm pion gas}

To illustrate how chiral symmetry constrains hadron properties in matter, I
consider first the case of matter at finite temperature in the chiral limit.
At low temperatures and zero baryon density, hadronic matter is just a gas of
weakly interacting, massless pions. This has been studied for some time
[2--12] and the results are cleaner than those for cold nuclear matter.

The scalar density of quarks in a pion in the chiral limit is, from
(\ref{sigpipi}),
\begin{equation}
\langle\pi|\overline\psi\psi|\pi\rangle=\left.{m_\pi^2\over\bar m}
\right|_{\bar m\rightarrow 0}=-{1\over f_\pi^2}\langle 0|\overline\psi\psi
|0\rangle,
\end{equation}
and the scalar density of pions in the gas, obtained by averaging over the 
Bose-Einstein distribution, is
\begin{equation}\label{sdpi}
\langle{\textstyle{1\over 2}}\mbox{\boldmath$\phi$}^2\rangle_T={T^2\over 8}.
\end{equation}
With these we can calculate the change in the quark condensate to order $T^2$:
\begin{equation}
\langle\overline\psi\psi\rangle\simeq\langle 0|\overline\psi\psi|0\rangle
+\langle{\textstyle{1\over 2}}\mbox{\boldmath$\phi$}^2\rangle_T\;
\langle\pi|\overline\psi\psi|\pi\rangle
=\langle 0|\overline\psi\psi|0\rangle\left(1-{1\over 8}{T^2\over f_\pi^2}
\right).
\end{equation}

In the presence of matter, the leading change in the mass of any particle is 
given by a sum over (the scalar parts of) the amplitudes for the scattering of 
the ``probe" particle from the various particles in the medium. For a heavy
hadron, such as a vector meson or nucleon, chiral symmetry requires that the
isoscalar hadron-pion scattering amplitude be of order $q^2$ (where $q$ is the
pion momentum) in the chiral limit. As a result the change in the mass of such
a particle is proportional to an integral over the Bose-Einstein distribution
of pions weighted with an extra factor of momentum-squared relative to
(\ref{sdpi}). Since the typical momenta of the pions in the gas are of order
$T$, the changes in hadron masses are of order $T^4$ instead of $T^2$. This
shows that hadron masses in the gas cannot scale like the the quark condensate
(or any simple function of $\langle \overline\psi\psi\rangle$).

In fact, by analogy with the behaviour of a superfluid~\cite{fw}, we should
not have expected any simple relation between these quantities. There the
condensate density (the order parameter) changes at order $T^2$ while the
superfluid density (which is a response function) changes at order $T^4$.
Hadron masses can be defined in terms of response functions of the QCD vacuum
and so behave in a similar way to the superfluid density, and quite
differently from the condensate.

One of the lessons to be learned from this is the changes in the quark
condensate from low-momentum pions do {\em not} necessarily signal partial
restoration of the symmetry. Large-amplitude, low-momentum fluctuations of the
pion fields around the chiral circle (or indeed a pion condensate) can
significantly reduce the average value of the condensate without moving the
system off the chiral circle. In such a case the hadron spectrum remains
unchanged and so the chiral symmetry is still hidden. One should remember that
$\langle\overline\psi\psi\rangle$ is only one possible order parameter; others
can become important if $\langle\overline\psi\psi\rangle$ is 
small.\footnote{The pion would then be anomalously light~\cite{cb}, with a mass
proportional to $\bar m$ rather than its square root. There is a school of
chiral perturbation theory that suggests that this is already the situation in
the normal vacuum~\cite{stern}.}

We can excite a superfluid just by pouring it. Since this action is invariant
under phase rotations of the condensate, any changes in the response function
reflect changes in the excitation spectrum. In contrast the easiest ways to
study the response of the QCD vacuum involve operators that are not chirally
invariant. As a result the corresponding response functions can contain
temperature-dependent pieces that reflect the chiral transformation properties
of the operators and have nothing to do with any changes in the spectrum. A
typical example is the correlator of two isovector vector currents,
\begin{equation}
C_{\mu\nu}^V(p,T)=(2\pi)^{-4}\int\!d^4x\, e^{ip\cdot x}\langle{\rm T}[V_\mu(x),
V_\nu(0)]\rangle_T.
\end{equation}

The pion gas changes the couplings of the currents to physical hadrons and 
mixes them with the axial currents~\cite{dei,ei1,ei2}, so that to leading order
in $T^2$ the correlator is
\begin{equation}\label{vcorrt}
C_{\mu\nu}^V(p,T)\simeq\left(1-{1\over 6}{T^2\over f_\pi^2}\right)
C_{\mu\nu}^V(p,0)+{1\over 6}{T^2\over f_\pi^2}C_{\mu\nu}^A(p,0),
\end{equation}
where, after averaging over isospin, the leading effects of the pion gas can be
expressed in terms of ${4\over 3}\langle\alpha|{1\over 2}\mbox{\boldmath
$\phi$}^2|\alpha\rangle$. The effect on the correlator of axial currents is
very similar, with the pion decay constant becoming~\cite{ggl}
\begin{equation}
f_\pi(T)=f_\pi\left(1-{1\over 12}{T^2\over f_\pi^2}\right).
\end{equation}

However, even though the couplings change and the correlators mix at order
$T^2$, the spectrum of states excited by the currents in (\ref{vcorrt}) is
still that of the zero-temperature correlators $C_{\mu,\nu}^{V,A}(p,0)$. Hence
to this order the $\rho$-$a_1$ splitting is unchanged and there is no
restoration of chiral symmetry in the spectrum.

These contributions to correlators in matter arising from the chiral
transformation properties of the operators mean that we need to be rather
careful when using QCD sum rules to study hadrons in matter. These sum rules
are obtained by using the operator product expansion (OPE) to relate
correlators to condensates (expectation values of various local operators).

A particularly simple case is the correlator of two isoscalar vector currents,
which can be used to derive a sum for the mass of the $\omega$ meson. The OPE
of this correlator involves only chirally invariant operators such as
$[\overline\psi \gamma^\mu(1\pm\gamma_5)\lambda_a\psi]^2$. The expectation
values of these are known as four-quark condensates. They are often estimated
by Fiertz rearranging to write them in terms of $\langle(\overline\psi
\psi)^2\rangle$ and then assuming a factorised form, $\kappa\langle
\overline\psi \psi\rangle^2$. However for a chirally invariant operator ${\cal
O}$ a soft-pion theorem gives
\begin{equation}
\langle\pi(q)|{\cal O}|\pi(q)\rangle\simeq-{1\over f_\pi^2}\langle 0|[Q_5,[Q_5,
{\cal O}]]|0\rangle=0,
\end{equation}
and so there should be no contributions to the condensates that are
proportional to the scalar density of pions. The leading temperature
dependence of the OPE representation of the correlator is thus of order $T^4$.
This matches with the order-$T^4$ change in the $\omega$ mass, but it is
not consistent with the factorised ansatz often used for the four-quark 
condensates.

The correlators used to derive sum rules for the $\rho$ meson and nucleon
masses are more complicated. The OPE's of these include operators that are not
chirally invariant and so give rise to terms of order $T^2$ but a careful
treatment of low-momentum pion terms in the spectral representations of the
correlators shows that these exactly cancel the pieces of order $T^2$ from the
OPE~\cite{elet1,elet2,ei1,ei2}. This leaves leading changes in the masses that
are of order $T^4$.

\section{Nuclear matter}

The discussion of nuclear matter is somewhat more complicated because the pions
involved are virtual rather than real, and because the nucleons are strongly 
interacting. Nonetheless the same basic features are present~\cite{birse}.

Let me start with the simple additive estimate of the change in the quark 
condensate in nuclear matter with a scalar density $\rho_s$ of 
nucleons~\cite{cfg}:
\begin{eqnarray}
\langle\overline\psi\psi\rangle_\rho&\simeq&\langle 0|\overline\psi\psi
|0\rangle+\rho_s\langle N|\overline\psi\psi|N\rangle\nonumber\\
&\simeq&\langle 0|\overline\psi\psi|0\rangle\left(1-{\rho_s
\sigma_{\pi{\scriptscriptstyle N}}\over f_\pi^2 m_\pi^2}\right).
\end{eqnarray}
This suggests a $\sim 30\%$ reduction in the quark condensate in nuclear
matter with a density of $\rho=0.17$ fm$^{-3}$.

This change in the condensate contains a significant contribution from 
low-momentum pions, in this case virtual particles in the pion clouds of the 
nucleons. Like the real pions in the gas, these cannot contribute to hadron
masses in matter. To pick out their contribution we can make use of the methods
of chiral perturbation theory (ChPT)~\cite{bkm}.

The relevant piece is proportional to the scalar density of pions in a nucleon.
The chiral expansion of this quantity in powers of $m_\pi$ has the form
\begin{equation}\label{pisdn}
\langle N|{1\over 2}\mbox{\boldmath$\phi$}^2|N\rangle\simeq A_\pi
-{9\over 16\pi}\left({g_{\pi{\scriptscriptstyle NN}}\over 2
M_{\scriptscriptstyle N}}\right)^2
m_\pi+\cdots.
\end{equation}
One can write the sigma commutator as a sum of core and cloud contributions,
\begin{eqnarray}
\sigma_{\pi{\scriptscriptstyle N}}&=&A_{\rm core}m_\pi^2+\langle N|{\textstyle
{1\over 2}}\mbox{\boldmath$\phi$}^2|N\rangle\langle\pi|\bar m\overline\psi\psi
|\pi\rangle\nonumber\\
&\simeq&\left[A-{9\over 16\pi}\left({g_{\pi 
{\scriptscriptstyle NN}}\over 2 M_{\scriptscriptstyle N}}\right)^2m_\pi\right]
m_\pi^2.
\end{eqnarray}
The constant $A\;(=\!A_{\rm core}+A_\pi)$ is a ``counterterm" in ChPT. It
contains both short-distance (core) and long-distance (cloud) contributions,
which cannot be separated in a model-independent way. In contrast the second
term, which has a nonanalytic dependence on $m_\pi^2$, is a purely
long-distance effect. It arises from the lowest-momentum pions in the tail of
the pion cloud. It is model-independent, with a coefficient given entirely in
terms of the $\pi N$ coupling and nucleon mass. Moreover, being a
long-distance effect, it is unaffected by short-range correlations between the
nucleons. We can therefore use nonanalytic terms like this as markers for the
contributions of low-momentum pions.

 From (\ref{pisdn}) we see that there is a contribution from these pions to the
quark condensate in matter that is of order $m_\pi\rho$. For hadron masses to
scale with $\langle\overline\psi\psi \rangle_\rho$, there would need to be a
term of order $m_\pi$ in the hadron-nucleon scattering amplitude. However
Weinberg's power counting shows that no such term in
present~\cite{wein}.\footnote{Strictly, Weinberg's counting rules apply to the
two-particle irreducible scattering amplitude. However this is the relevant
amplitude for the definition of a mass that could appear in, for example, a
Dirac equation for a nucleon in matter. It avoids the problem of strong energy
dependence of the full scattering amplitude produced by poles close to
threshold.} Instead the leading nonanalytic contribution from isoscalar
two-pion exchange to hadron-nucleon scattering is of order $m_\pi^3$, which
gives rise to a leading change in the hadron mass of order $m_\pi^3\rho$. This
suppression by two chiral powers relative to the scalar density of pions has
exactly the same origin as in the pion gas: the leading terms of the isoscalar
pion-hadron scattering amplitude are of order $q^2$ and $m_\pi$, as required
by chiral symmetry. Hence we again see that hadron masses cannot scale like
any simple function of $\langle\overline\psi\psi \rangle_\rho$ in
matter~\cite{birse}.

There are also similar contributions from low-momentum pions to the couplings
of vector and axial currents to hadrons. These can lead to changes in
quantities like $f_\pi$ and $g_A$ which are proportional to $\langle{1\over
2}\mbox{\boldmath$\phi$}^2 \rangle$ in isospin symmetric nuclear matter. The
latter can show up as a two-pion exchange contribution to the quenching of
$g_A$~\cite{cde}. There is also an analogue of the mixing of vector and axial
correlators~\cite{krippa} which can, in principle, affect pion
photoproduction~\cite{cde}.

\section{Conclusions}

We have seen that there are no simple relations between changes to the quark
condensate, the pion decay constant and hadron masses in matter. Specifically,
low-momentum real or virtual pions generate terms in the condensate of order
$T^2$ in a pion gas or $m_\pi\rho$ in nuclear matter that are absent from the
masses. As a result, hadron masses cannot scale like any simple function of the
condensate (or of $f_\pi$) in matter. This is a consequence of the chiral
suppression of the interactions between low-momentum pions and other hadrons,
which means that contributions to the condensate from the scalar density of
pions cannot lead to changes in the masses of heavy hadrons like vector mesons
and nucleons.

Operators like the vector and axial currents are not chirally invariant and so
low-momentum pions can contribute to observables like $f_\pi$ and $g_A$. In
isospin symmetric matter, the lowest order pieces of the changes in such
quantities can be expressed in terms of the scalar density of pions. However
the operators have different isospin structures from the quark condensate and
so these terms have different coefficients compared with the similar term in
$\langle\overline\psi\psi\rangle$. More generally there is no simple relation
between such couplings and the condensate in matter.

These results show the importance of including pionic fluctuations in any
study of hadron properties in matter.\footnote{See also the contributions of
M.~Ericson, B.~Friman, V.~Koch, J.~Wambach and W.~Weise to this workshop.}
Pionic fluctuations can change the average value of the quark condensate with
moving the system off the chiral circle, and so a decrease in one order
parameter, such as the condensate, on its own need not be a signal of partial
symmetry restoration. One needs to look at the response functions of the
vacuum as well, and in particular the hadron spectrum.

Let me finish with a question: Can we find some other order parameter or
similar quantity with a more direct relation to hadron masses than the quark
condensate?

\section*{Acknowledgements}

I am grateful to B. Krippa and J. McGovern for helpful discussions. This work
was supported by the EPSRC.

\bibliographystyle{unsrt}

\end{document}